# Characteristic of Paddle Squeezing Angle and AMBU Bag Air Volume in Bag Valve Mask Ventilator

Cong Toai Truong[3], Kim Hieu Huynh[3], Van Tu Duong[1,2,3,*], Huy Hung Nguyen[3,4], Le An Pham[5], and Tan Tien Nguyen[1,2,3,*]

[1]*Faculty of Mechanical Engineering, Ho Chi Minh City University of Technology (HCMUT), 268 Ly Thuong Kiet, District 10, Ho Chi Minh City, Vietnam*

[2]*Vietnam National University Ho Chi Minh City, Linh Trung Ward, Thu Duc District, Ho Chi Minh City, Vietnam*

[3]*National Key Laboratory of Digital Control and System Engineering (DCSELab), HCMUT, 268 Ly Thuong Kiet, District 10, Ho Chi Minh City, Vietnam*

[4]*Faculty of Electronics and Telecommunication, Saigon University, Vietnam*

[5]*Grant and Innovation Center (GIC), University of Medicine and Pharmacy at Ho Chi Minh City*

*Corresponding author: dvtu@hcmut.edu.vn, nttien@hcmut.edu.vn

## Abstract

In the COVID-19 period, the number of deaths has increased every day around the world. The pandemic has impacted the life and economy. Especially, there is a shortage in medical including a lack of technology, facility and equipment. One of those, ventilators are the essential equipment that does not provide enough requirements for the hospital. A ventilator is an essential unit in hospitals because it seems to be the first step to protect the life of the patient getting sick. Some low-income countries aim to make a simple ventilator using locally available and low- cost materials for primary care and palliative care. One of the simple principles of ventilators is to adopt an artificial manual breath unit (AMBU) bag with paddles. Unfortunately, the squeezing angle of paddles is not proportional to the exhaust air volume from the AMBU bag. This paper analyzes the character of the squeezing angle of the paddles and the exhaust air volume of the adult AMBU bag through experiments. The result can be used to control the squeezing angle through a DC motor mounted with paddles to obtain the desired air volume.

## 1 Introduction

The coronavirus (COVID-19) pandemic is very complicating and multi-staged manner. Currently, COVID-19 has many dangerous strains and has not had the signal of decreasing around the world. Besides that, the COVID-19 patients must be treated with ventilators which have many higher functional requirements in Intensive Care Unit, but also need a simple assist ventilator in primary, palliative care, and safety transportation. Thus, the number of modern-ventilator is a big challenge for the health (Ercole *et al.*, 2009; P Smetanin, D Stiff, A Kumar, P Kobak, R Zarychanski, N Simonsen, 2009; Stiff *et al.*, 2011; Wiederhold and Riva, 2013), especially in developing countries (Fisher and Heymann, 2020; The Lancet, 2020; WHO, 2020). Several attempts have been made to create low-cost ventilators (Russell and Slutsky, 1999; Fang *et al.*, 2020). In particular, there are machines operating using the grippers (Kwon *et al.*, 2020) to deal with overloading COVID-19 cases. Medically, although some studies showed that the 2-handed mask-face technique is prone to be better than 1-handed mask-face technique (Jesudian *et al.*, 1985; Wheatley *et al.*, 1997; Davidovic, LaCovey and Pitetti, 2005; Joffe, Hetzel and Liew, 2010), in some emergency situations, the 1-handed technique is still applied due to the lack of AMBU compression which can be conducted by the ventilator using the grippers. Therefore, this type of ventilator is very essential which has the advantages of easy manufacturing and assembly, low cost, easy-to-find materials and control volume adjustment according to the angle of the paddles. However, the biggest difficulty when developing the BVM ventilator is that it is quite difficult to identify the mathematical model of the AMBU bag making the control process becomes complicated. To be more specific, during the development the BVM ventilator, the most important device is the AMBU bag (not only standard in ER, but also very popular in primary care), defining the characteristic correlation between the squeezing angle of the ventilator and the volume exiting the AMBU bag contributes to the development of controllers that allows the ventilator to achieve better controlling performance in volume-controlled mode. Therefore, this paper studies on the BVM ventilator in order to investigate the characteristic of the grippers angle of the BVM ventilator and the air volume exiting the AMBU bag. To achieve the result, the experiment was conducted by controlling the gripper to rotate at a random angle in each cycle, within the limitation of the paddles' angle before reaching the safety limit switch. The obtained data includes the squeezing angle measured from the motor encoder and the volume exiting the AMBU bag calculated from the flow value read from the flow sensor. The flow sensor has been calibrated to zero in normal condition, and the air volume calculation is conducted by integrating the flow value slices along time from the beginning of the inhalation to the beginning of the exhalation. The interpolation function is determined from the least squares regression method and the acceptability of the interpolation function is evaluated from the correlation coefficient, root mean square error, sum square error and mean absolute error value. Through analyzing and building characteristics based on simulation and experiment, this research find out the relationship between the paddles and the AMBU bag (for adults), and the mathematical model of the relationship between grippers angle and exhausted air volume is obtained.

## 2  Materials and Methods

### 2.1.  Overall system description

Breath circuit is depicted in Figure 1 with the usage of bag valve mask ventilation system. At the inspiration phase, two paddles squeeze the artificial manual breath unit (AMBU) bag which allows oxygen to move from the air mixer through the humidifier and the HEPA filter into the patient's lungs. Following this is the spontaneous exhaling of the patient into the external environment due to the elasticity of the lungs and the operation of the one-way valve. The positive end-expiratory pressure (PEEP) valve should be placed closed to the patient wearing a mask in order to avoid the exceeding pressure in the lung. The working principle of PEEP van is that the patient exhales against a spring, and PEEP valve is used to keep up the pressure higher than an adjustable value that keeps the alveoli from collapsing in the expiration phase. Air mixture is humidified and warmed before ventilating by the humidifier. HEPA filter is used for removing harmful elements in the air such as dust, pet dander, etc. Besides, for the purpose of avoiding barotrauma, a relief valve is placed close to the patient which prevents overpressure during the breath.

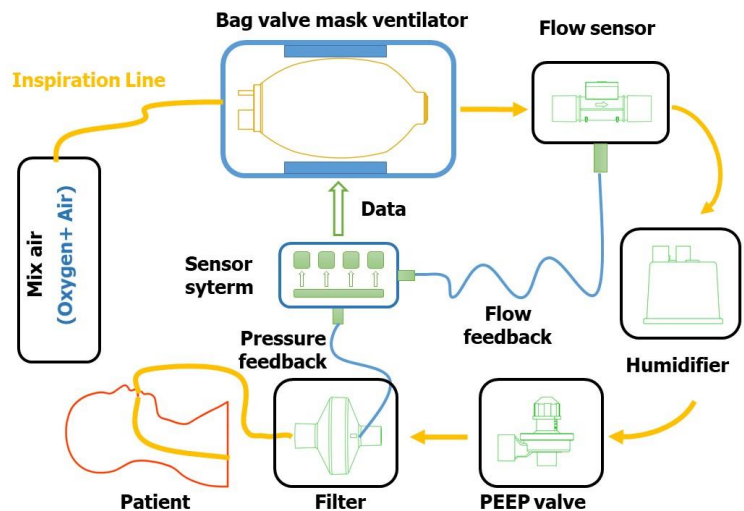

**Figure 1. The operating system of bag valve mask ventilator.**

### 2.2.  Bag Valve Mask Ventilator

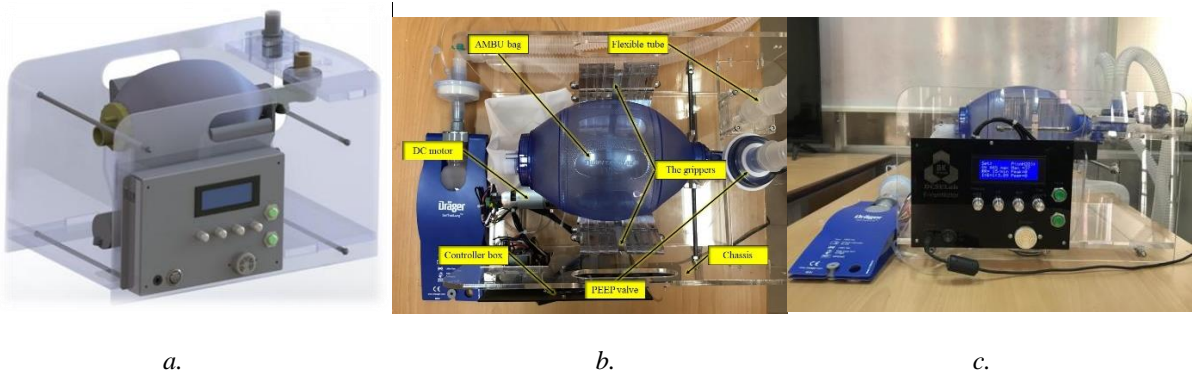

*a.* *b.* *c.*

**Figure 2. The modeling to show the relationship between the grippers and AMBU bag.**

Figure 2a shows a 3D model of a ventilator operating based on BVM system. Figures 2 b, c shows the experimental model of the machine with the listed components including: AMBU bag, flexible tube, PEEP valve, the grippers, DC motor, and controller box. The operation of the BVM ventilator is started by pressing the Start/Home button in order to return the two clamps to its home position by sensing the signal from a limit switch. After the homing process, the clinicians can vary the parameters displayed on the ventilator's LCD including $V_T, I:E, RR, PIP$. In addition, there are two other parameters displayed on the ventilator, the PEEP value and the current pressure on the breath circuit. After the initial setting, the clinicians can press the Start/Home button again to confirm the parameter change and allow for the machine to operate. Following this is the squeezing of the two paddles at regular cycle. The BVM might generate a warning in the following cases: the current pressure exceeds the PEEP and PIP limit range; the doctor set the parameters without pressing Start/Home button to confirm; the squeezing angle of the paddles exceeds the limitation large and reaches the safety limit switch.

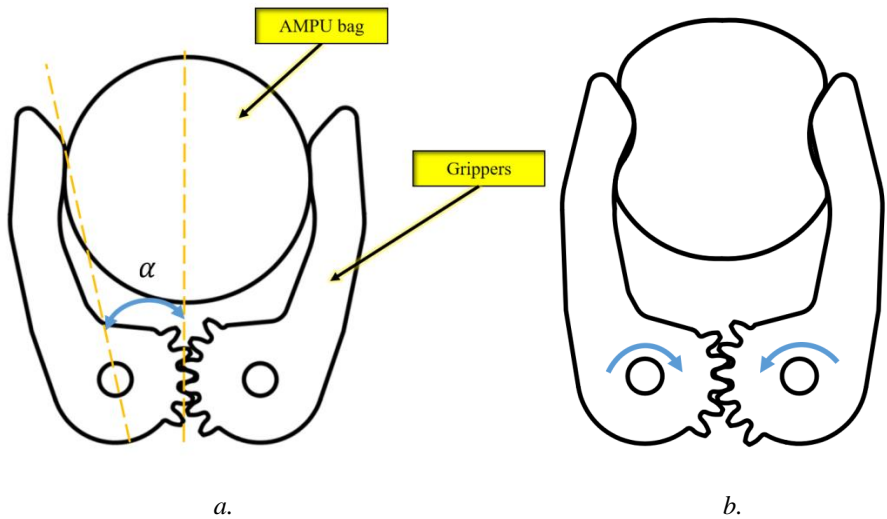

*a.* *b.*

**Figure 3. Determination between grippers and AMBU bag.**

Figure 3 shows the angle ($\alpha$) and grippers of projection directly in the front view of the BVM. Figure 3a shows the state of grippers and AMBU bag in the homing position. At this initial position, the gripper touches the AMBU bag. In this position, $\alpha$ is conceptualized as the initial angle of the system, which is adjusted based on the position sensor to ensure the BVM start position. When the BVM is operating, the motor providing torque to the grippers by the gear transmission. The grippers of BVM impact the AMBU bag causing the bag to deform without a fixed shape as shown in Figure 3b. Therefore, the squeezing angle causes the exhausted air volume to change nonlinear which make it difficult to determine the tidal volume of air traveling into the patient's lungs. In particular, with the silicon mechanical properties of AMBU bags determining the exhausted air volume more difficult. Thus, to solve the aforementioned issue, the calculation of the $\alpha$ and the volume of the AMBU bags are carried out as a basis for the controller studies in the next section.

### 2.3. Theoretical basis

The AMBU bag adopted for this research is adult unit. Thus, The specification on the AMBU bag is referred to (UW Health, 2019) and shown in table I.

Table I. Manual-Resuscitator of AMBU Adult

| Body weight | > 40kg |
|---|---|
| Stoke volume | 800 ml |
| Resuscitator volume | 1650 ml |
| Dimensions (Length x Diameter) | 212x131 mm |
| Bag reservoir volume | 2000 ml |

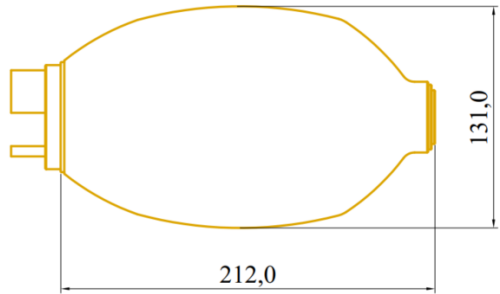

**Figure 4. Dimension 2D of AMBU bag.**

Due to the circular shape, the volume of the AMBU bag is approximated by the following integral:

$$V = \pi \int_{\alpha}^{\beta} f(x)^2 dx \qquad (1)$$

Where, $(\alpha, \beta)$ = [-106;106]; $f(x)$ is the AMBU contour curve equation. According to Table I, the AMBU's volume provided by the manufacturer is equal to $1650 ml$. Eq. (1) becomes:

$$\pi \int_{\alpha}^{\beta} [f(x)]^2 dx = 1650 \qquad (2)$$

$$\Leftrightarrow \int_{-110}^{110} [f(x)]^2 dx = \frac{1650}{\pi} \qquad (3)$$

Solving Eq. ((3)) for $f(x)$ is very complicated, therefore, we approximate $f(x)$ as a quadratic function satisfying the following conditions (Otten *et al.*, 2014) :

$$f(x = -106) = f(x = 106) = 0 \tag{4}$$

$$f(x = 0) = 65.5 \tag{5}$$

$$f(x) = ax^2 + bx + c \tag{6}$$

By substituting Eq. (4) – (5) into Eq.(6)(3), $f(x)$ becomes:

$$f(x) = -0.00583x^2 + 65.5 \tag{7}$$

Substitute Eq. (7)(8) into Eq. (3), it yields:

$$V = \pi \int_{\alpha}^{\beta} [f(x)]^2 dx = \frac{\pi}{1000} \int_{-106}^{106} (-0.00583x^2 + 65.5)^2 dx \approx 1523.87 \ (ml) \tag{8}$$

The approximate contour equation of the AMBU bag gives the bag's volume error of 7.6% compared to the volume value provided by the manufacturer.

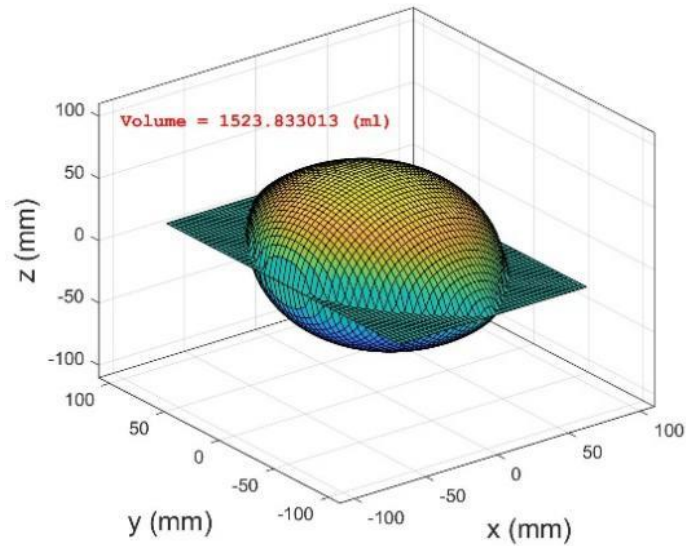

**Figure 5. Simulation of AMBU bag by linear interpolation method.**

When it comes to simulation, the volume result of AMBU bag calculated by linear interpolation is approximately $1523.87 \ ml$. Approximating the squeezing process with the exhaust air from the AMBU bag being outside the $x = -d$ and $x = d$ planes. Assuming $d = 30mm$, the remaining volume of the AMBU bag is calculated by the following simulation.

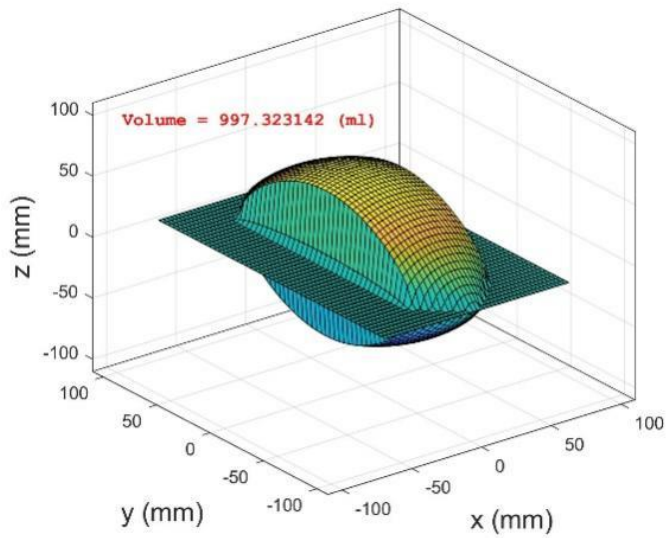

**Figure 6. Simulation of AMBU bag at $d = 30mm$.**

According to Figure 6, the remaining air volume in the AMBU bag is $\approx 997.32\ (ml)$, the exhaust air volume is therefore calculated as $\approx 526.51\ (ml)$. Given the increasing of $\alpha$ from $0^o$ (home position) to $32^o$, we determine the graph of the characteristic of the grippers angle $\alpha$ and the exhaust air volume $V_T$:

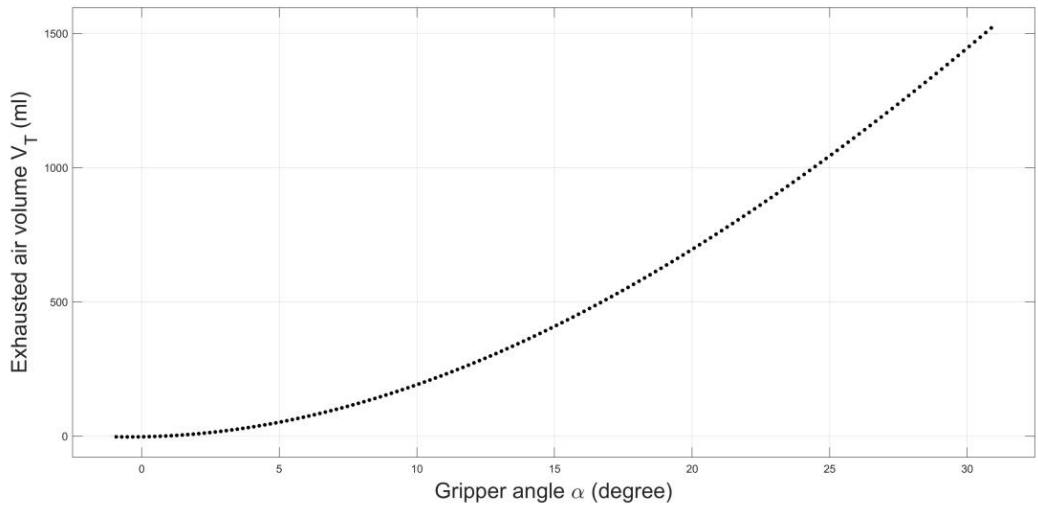

**Figure 7. The characteristic of grippers angle and exhaust air volume from AMBU bag.**

The characteristic showed in Figure 7 is determined to be nonlinear. However, within the operation of the BVM ventilator, $V_T$ has a range from $350ml$ to $700ml$. Figure 8 display the characteristic showed in Figure 7 with $V_T$ being within the required range.

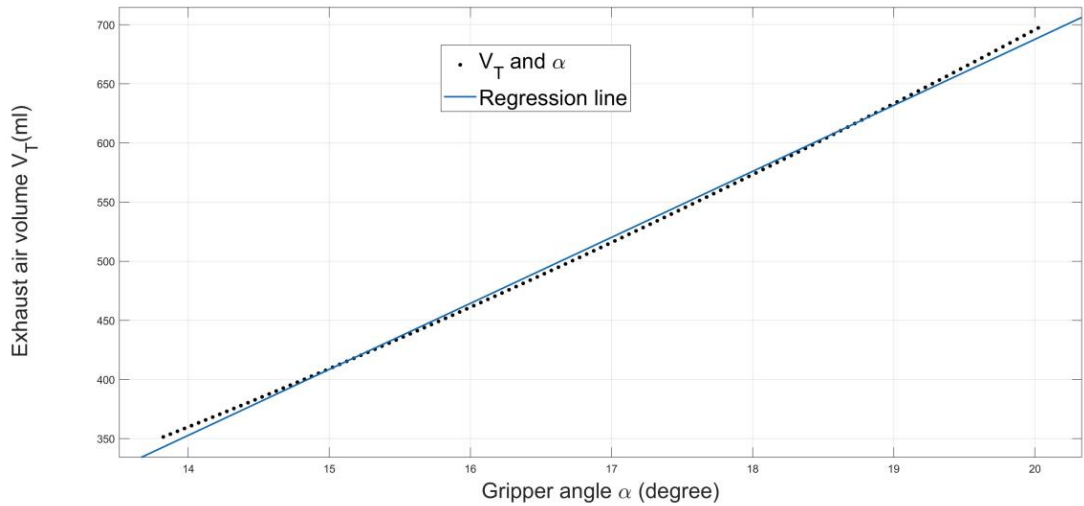

**Figure 8. The characteristic of grippers angle and exhaust air volume from AMBU bag within the exhausted air volume range from 350ml to 700ml.**

The theoretical characteristic between grippers angle $\alpha(degree)$ and exhaust air volume $V_T(ml)$ can be approximated as linear with the correlation coefficient $R \approx 0.9985$. The equation shows the relationship can be written as follow:

$$V_T(t) = 55.81 \times \alpha(t) - 428.6 \tag{9}$$

## 3 Experimental Result

To evaluate the theory, an experiment on the actual BVM ventilator model was conducted, the obtained data including grippers angle calculated from motor encoder and exhaust air volume calculated from flow sensor.

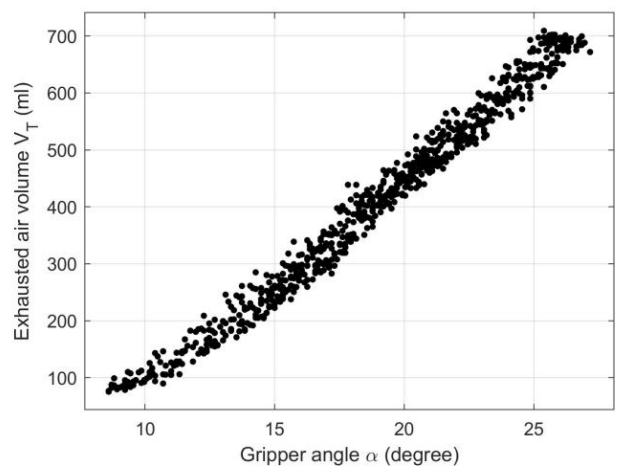

**Figure 9. The experimental characteristic of grippers angle and exhaust air volume from AMBU bag.**

The nonlinearity of the characteristic shown in Figure 9 has been identified. It can be seen that the shape of the theoretical characteristic curve presented in Figure 7 is similar to the experimental one in Figure 9. Figure 10 shows the same feature as Figure 9, but with $V_T$ inside the the operating range of the BVM ventilator ($350ml \div 700ml$).

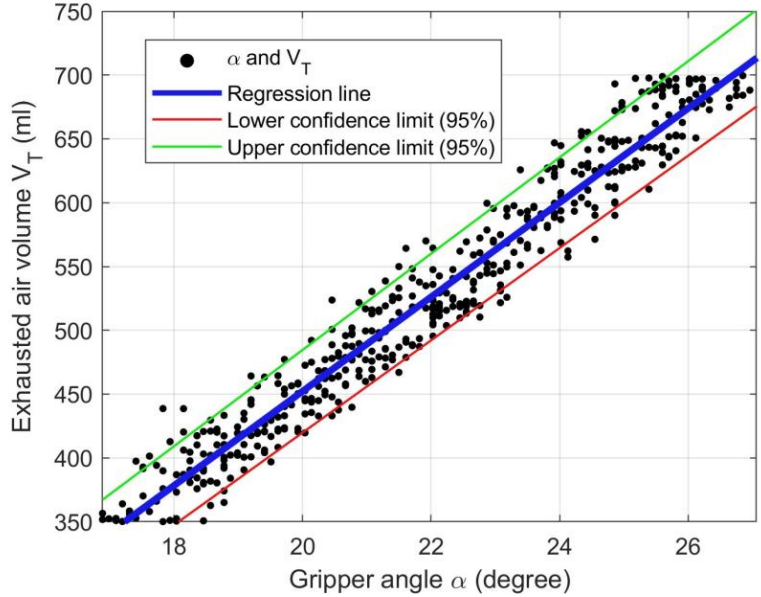

**Figure 10. The experimental characteristic of grippers angle and exhaust air volume from AMBU bag with the exhaust air volume range from $350ml$ to $700ml$.**

The experimental characteristic of the two variables $V_T$ and $\alpha$ can be approximated as a linear function with the fitting properties presented in Table II.

Table II. Regression properties of experimental characteristic presented in Figure 10

| | |
|---|---|
| Linear model Poly1 | $V_T(t) = p_1 \times \alpha(t) + p_2$ |
| Coefficients (95% confidence bounds) | $p_1 = 36.96\ (36.19, 37.73)$ |
| | $p_2 = -287.1\ (-304.1, -270.1)$ |
| Goodness of fit: | Sum square error (SSE): $22900(ml^2)$ |
| | Coefficient of determination (R-square): $0.9505$ |
| | Adjusted R-square: $0.9504$ |
| | Root mean square error (RMSE): $22.17(ml)$ |

The interpolation function has coefficient of determination of $0.9505$ The sum square error is $22900(ml^2)$ and root-mean-square error of $22.17(ml)$. In general, the actual and the theoretical equation has the same form of function, but the coefficients are different because the AMBU bag is dilated during squeezing which make the exhausted air volume in the simulated model smaller than the actual model. In other word, the exhaust air volume AMBU in the simulation model is lower than that of the actual model with the same value of grippers angle. In short, $V_T$ is approximated by two components, the linear component $\alpha(t)$ with slope $p_1 = 36.36$ and the constant component with the value of $p_2 = -287.1$.

$$V_T(t) = V_{T,1} + V_{T,2} = 36.96 \times \alpha(t) - 287.1 \tag{10}$$

In order to evaluate the regression result of the Eq. (10), another experiment was conducted. The scenario of the experiment consists of carrying out several times squeezing with fixed input gripper angles, the output exhausted air volume is obtained. The evaluation result is shown in Figure 11.

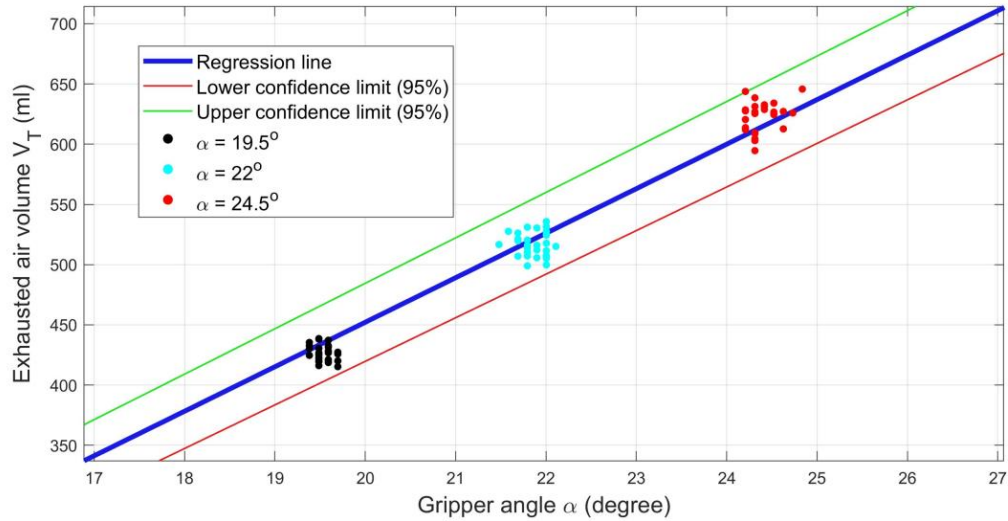

**Figure 11. Evaluation of the regression line presented in Figure 10**

It can be seen that the obtained exhausted air volumes are within the confidence limit lines. In a more clear way, at $\alpha$ equal to $19.5^o, 22^o, 24.5^o$ respectively, we calculate root mean square error values (RMSE) and mean absolute error values (MAE) of the obtained exhausted air volume with respect to volume values calculated from the regression line. The results are shown in Table III in purpose of evaluating the operating error of the ventilator compared to the regression line.

Table III. Operating error of the ventilator at fixed gripper angles compared to the regression line.

| Gripper angle | RMSE | MAE |
|---|---|---|
| $\alpha = 19.5^o$ | 9.9138 $(ml)$ | 6.1409 $(ml)$ |
| $\alpha = 22.5^o$ | 11.6461 $(ml)$ | 9.6960 $(ml)$ |
| $\alpha = 24.5^o$ | 13.9763 $(ml)$ | 10.1799 $(ml)$ |

# 4 Conclusions

As aforementioned in Section 1, the correlation between the AMBU bag and the grippers angle is the most critical factor of the BVM ventilator since the ventilation results depends mainly on the grippers' controlling results and the behaviors of the AMBU bag. Thus, determining the characteristic between the ventilator's gripper angle and the exhausted air volume of the AMBU bag plays an important role in improving the BVM ventilator controlling performance. Our research could have potential applications in the future when it comes to the production and the control of ventilators based on the BVM system. It can be seen that there were errors in the BVM ventilator modeling process. More specific, although the coefficients of the simulated correlation equation between the gripper angle and the exhausted air volume are different compared to the experimental correlation, the simulated correlation equation and the experimental correlation equation still have the same form within the confidence limit. In short, based on the presented BVM ventilator modeling process, the result of this research make effort to reduce errors in future studies.

## Acknowledgments

This research is supported by DCSELAB and funded by Vietnam National University Ho Chi Minh City (VNU-HCM) under grant number TX2021-20b-01. We acknowledge the support of time and facilities from Ho Chi Minh City University of Technology (HCMUT), VNU-HCM for this study. This research is also funded by Department of Science and Technology under grant number 58/2020/HĐ-QPTKHCN.s

## References

Davidovic, L., LaCovey, D. and Pitetti, R. D. (2005) 'Comparison of 1- versus 2-person bag-valve-mask techniques for manikin ventilation of infants and children', *Annals of Emergency Medicine*, 46(1), pp. 37–42. doi: 10.1016/j.annemergmed.2005.02.005.

Ercole, A. *et al.* (2009) 'Modelling the impact of an influenza A/H1N1 pandemic on critical care demand from early pathogenicity data: The case for sentinel reporting', *Anaesthesia*, 64(9), pp. 937–941. doi: 10.1111/j.1365-2044.2009.06070.x.

Fang, Z. *et al.* (2020) 'AmbuBox: A Fast-Deployable Low-Cost Ventilator for COVID-19 Emergent Care', *SLAS Technology*, 25(6), pp. 573–584. doi: 10.1177/2472630320953801.

Fisher, D. and Heymann, D. (2020) 'Q&A: The novel coronavirus outbreak causing COVID-19', *BMC Medicine*, 18(1), pp. 18–20. doi: 10.1186/s12916-020-01533-w.

Jesudian, M. C. S. *et al.* (1985) 'Bag-valve-mask ventilation; two rescuers are better than one: Preliminary report', *Critical Care Medicine*, pp. 122–123. doi: 10.1097/00003246-198502000-00015.

Joffe, A. M., Hetzel, S. and Liew, E. C. (2010) 'A two-handed jaw-thrust technique is superior to the one-handed "EC-clamp" technique for mask ventilation in the apneic unconscious person', *Anesthesiology*, 113(4), pp. 873–879. doi: 10.1097/ALN.0b013e3181ec6414.

Kwon, A. H. *et al.* (2020) 'Rapidly scalable mechanical ventilator for the COVID-19 pandemic', *Intensive Care Medicine*, 46(8), pp. 1642–1644. doi: 10.1007/s00134-020-06113-3.

Otten, D. *et al.* (2014) 'Comparison of bag-valve-mask hand-sealing techniques in a simulated model', *Annals of Emergency Medicine*, 63(1), pp. 6-12.e3. doi: 10.1016/j.annemergmed.2013.07.014.

P Smetanin, D Stiff, A Kumar, P Kobak, R Zarychanski, N Simonsen, F. P. (2009) 'Potential intensive care unit ventilator demand / Canada', *Journal of Infectious Diseases, The*, 20(4), pp. 115–123.

Russell, J. and Slutsky, A. (1999) 'International consensus conferences in intensive care medicine: Ventilator-associated lung injury in ARDS', *American Journal of Respiratory and Critical Care Medicine*, 160(6), pp. 2118–2124. doi: 10.1164/ajrccm.160.6.ats16060.

Stiff, D. *et al.* (2011) 'Potential pediatric intensive care unit demand/capacity mismatch due to novel pH1N1 in Canada', *Pediatric Critical Care Medicine*, 12(2), pp. 51–57. doi: 10.1097/PCC.0b013e3181e2a4fe.

The Lancet (2020) 'COVID-19: too little, too late?', *The Lancet*, 395(10226), p. 755. doi: 10.1016/S0140-6736(20)30522-5.

UW Health (2019) 'Manual Resuscitator Bag', *Department of Nursing*. Available at: https://www.uwhealth.org/healthfacts/respiratory/7820.pdf.

Wheatley, S. *et al.* (1997) 'A comparison of three methods of bag valve mask ventilation', *Resuscitation*, 33(3), pp. 207–210. doi: 10.1016/S0300-9572(96)01024-6.

WHO (2020) 'Critical preparedness, readiness and response actions for COVID-19: WHO/2019-nCoV/Community_Actions/2020.3', (March), pp. 1–3. Available at: https://www.who.int/publications-detail/critical-preparedness-readiness-and-response-actions-for-covid-19.

Wiederhold, B. K. and Riva, G. (2013) 'Original research', *Annual Review of CyberTherapy and Telemedicine*, 11, p. 63. doi: 10.1097/01.naj.0000529715.93343.b0.